\pdfoutput=1
\documentclass[twoside,leqno,twocolumn]{article}
\usepackage{ltexpprt}
\usepackage{graphicx}
\usepackage{cite}
\usepackage{hyperref}
\usepackage{url}

\graphicspath{{figures/}}

\begin{document}

\title{\huge\textsf{\textbf{Quadratic Time Algorithms Appear to be Optimal for Sorting Evolving Data}}}

\date{}

\author{
Juan Jos\'e~Besa \\[5pt]
Dept. of Computer Science \\
Univ. of California, Irvine \\
Irvine, CA 92697 USA \\
\texttt{jjbesavi@uci.edu}
\and
William E. Devanny\\[5pt]
Dept. of Computer Science \\
Univ. of California, Irvine \\
Irvine, CA 92697 USA \\
\texttt{wdevanny@uci.edu}
\and
David Eppstein \\[5pt]
Dept. of Computer Science \\
Univ. of California, Irvine \\
Irvine, CA 92697 USA \\
\texttt{eppstein@uci.edu}
\and
Michael T.~Goodrich \\[5pt]
Dept. of Computer Science \\
Univ. of California, Irvine \\
Irvine, CA 92697 USA \\
\texttt{goodrich@uci.edu}
\and
Timothy Johnson \\[5pt]
Dept. of Computer Science \\
Univ. of California, Irvine \\
Irvine, CA 92697 USA \\
\texttt{tujohnso@uci.edu}
}

\maketitle

\fancyfoot[R]{\footnotesize{\textbf{Copyright \textcopyright\ 2018 by SIAM\\
Unauthorized reproduction of this article is prohibited}}}

\begin{abstract}
We empirically study sorting in the \emph{evolving data} model.
In this model, a sorting algorithm maintains an approximation to the sorted
order of a list of data items while simultaneously, with 
each comparison made by the algorithm, an adversary randomly swaps the 
order of adjacent items in the true sorted order.
Previous work studies only two versions of quicksort, and has a gap between the
lower bound of $\Omega(n)$ and the best upper bound of $O(n \log \log n)$.
The experiments we perform in this paper provide empirical evidence that some
quadratic-time algorithms such as insertion sort and bubble sort are asymptotically
optimal for any constant rate of random swaps. 
In fact, these algorithms perform as well as or better 
than algorithms such as quicksort that are more efficient in the traditional algorithm
analysis model.
\end{abstract}

\section{Introduction}

In the traditional Knuthian model~\cite{knuth1998art}, an algorithm
takes an input, runs for some amount of time, and produces an output. 
Characterizing an algorithm in this model typically involves 
providing a function, $f(n)$, such that the running time
of the algorithm can be bounded asymptotically as being $O(f(n))$ on average,
with high probability (w.h.p.), or in the worst case.
Although this has proven to be an extremely useful model for general algorithm 
design and analysis, there are nevertheless 
interesting scenarios where it doesn't apply.

\subsection{The Evolving Data Model}
One scenario where the Knuthian model doesn't apply is
for applications where the input data is changing while an algorithm
is processing it, which has given rise to
the \emph{evolving data} model~\cite{sort11}. 
In this model, input changes are coming so
fast that it is hard for an algorithm to keep up, much like Lucy
in the classic conveyor belt 
scene \footnote{E.g., see \url{https://www.youtube.com/watch?v=8NPzLBSBzPI}.}
in the TV show, \textit{I Love Lucy}.
Thus, rather than produce a single output, as in the Knuthian model,
an algorithm in the evolving data model dynamically maintains 
an output instance over time.
The goal of an algorithm in this model 
is to efficiently maintain its output
instance so that it is ``close'' to the true output at that time.
Therefore, analyzing an algorithm in the evolving data model involves
defining a distance metric for output instances, parameterizing how 
input \emph{mutations} occur over time, 
and characterizing how well an algorithm
can maintain its output instance 
relative to this distance and these parameters for various 
types of input mutations.

% With respect to
% the sorting problem in this model,
% which is an original area of study for
% Anagnostopoulos~{\it et al.}~\cite{sort11},
% the underlying input is a set of 
% elements that belong to a total order that is mutating over time. 
The goal of a sorting algorithm in the evolving data model, then,
is to maintain an output order close to the true total order
even as it is mutating.
For example, the list could be 
an ordering of political candidates, tennis players, 
movies, songs, or websites, which is changing as the 
ranking of these items is evolving, e.g., see~\cite{RSSC:RSSC159}.
In such contexts, performing a comparison of two elements is
considered somewhat slow and expensive, in that it might involve a debate
between two candidates, a match between two players, or an online
survey or A/B test~\cite{dixon2011b} comparing two movies, songs, or websites.
In this model,
a comparison-based sorting algorithm would therefore be executing at
a rate commensurate with the speed in which the total order is changing, i.e.,
its mutation rate.
Formally, 
to model this phenomenon,
each time an algorithm performs a comparison, we allow for an
adversary to perform some changes to the true ordering of the elements. 

There are several different 
adversaries one might consider with respect to mutations that would
be anticipated after a sorting algorithm performs a comparison.
For example, an adversary (who we call the \emph{uniform} adversary) 
might choose $r>0$ consecutive pairs of elements in the true
total order uniformly at random and swap their relative order.
Indeed, previous work~\cite{sort11} provided theoretical analyses for 
the uniform adversary for the case when $r$ is a constant.
Another type of adversary (who we call the \emph{hot spot}
adversary) might choose an element, $x$, in the total order and 
repeatedly swap $x$ with a predecessor or successor each time 
a random ``coin flip'' comes up ``tails,'' 
not stopping until the coin comes up ``heads.''

A natural distance metric to use in this context is the \emph{Kendall tau}
distance~\cite{kendall1938new}, which counts the number of inversions between a total
ordering of $n$ distinct elements and a given list of those elements.
That is, a natural goal of a sorting algorithm in the evolving data
model is to minimize the Kendall tau distance for its output list
over time.

% In the evolving data framework, the input data changes during the execution of an algorithm. So instead of producing a single output, evolving data algorithms attempt to maintain an output close to the correct output for the current state of the data. At each time step, the algorithm is permitted a small interaction with the input data and then the input data undergoes a small \emph{mutation}.

% The exact evolving data model used is determined by the problem being studied, what constitutes an algorithm step, how the data mutates after each step, and how to judge the quality of the algorithm's answer.

% If the algorithm is particularly slow or the data is particularly volatile, then it is possible the input might be out of date and the output may no longer have relevance.
% For this reason, Anagnostopoulos~{\it et al.}~\cite{sort11} introduced the \emph{evolving data} framework.

Here, we consider the empirical performance of sorting
algorithms in the evolving data model. 
Whereas previous work looked only at quicksort
with respect to theoretical analyses against 
the uniform adversary, we are
interested in this paper in the ``real-world'' performance of a variety of sorting algorithms
with respect to multiple types of adversaries in the evolving data model.
Of particular interest are any experimental results 
that might be counter-intuitive or would highlight gaps in the theory.

\subsection{Previous Work on Evolving Data}

The evolving data model was introduced by
Anagnostopoulos~{\it et al.}~\cite{sort11},
who study sorting and selection problems with respect to an evolving
total order.
They prove that quicksort maintains a Kendall tau distance of 
$O(n\log n)$ w.h.p.~with respect to the true total order,
against a uniform adversary that performs a 
small constant number of
random swaps for every comparison made by the algorithm.
Furthermore, they show that a batched collection of quicksort
algorithms operating on overlapping blocks can maintain a Kendall tau
distance of $O(n\log\log n)$ against this same adversary. 
We are not aware of previous experimental work on sorting in the
evoloving data model.
We are also not aware of previous work, in general, in the evolving
data model for other sorting
algorithms or for other types of adversaries.

In addition to this work on sorting,
several papers have considered other problems in the evolving data
model. Kanade {\it et al.}~\cite{kanade_et_al:LIPIcs:2016}
study stable matching with evolving preferences. Huang {\it et
al.}~\cite{Huang2017} present how to select the top-$k$ elements
from an evolving list. Zhang~and~Li~\cite{zhang2016shortest}
consider how to find shortest paths in an evolving graph.
Anagnostopoulos~{\it et al.}~\cite{Anagnostopoulos:2012:AEG} study
$(s,t)$-connectivity and minimum spanning trees in evolving graphs.
Bahmani~{\it et al.}~\cite{Bahmani:2012} give several PageRank algorithms for
evolving graphs
and they analyze these algorithms both theoretically and experimentally. 
To our knowledge,
this work is the only previous experimental work for 
the evolving data model.

\subsection{Our Results}

% When studying a specific problem, a specific evolving data model is developed and studied. The specific evolving data model may not match precisely how the data in the real world application evolves.
% The theoretical proofs backing the performance of evolving sorting algorithms are very specific to the exact version of the evolving data model chosen.

In this paper, we provide an experimental investigation of sorting 
in the evolving data model.
The starting point for our work is 
the previous theoretical work~\cite{sort11} 
on sorting in the evolving data model, which only studies quicksort.
Thus, our first result is to report on experiments that address
whether these previous theoretical 
results actually reflect real-world performance.

In addition, we experimentally investigate a number of other classic 
sorting algorithms to empirically study whether these algorithms lead to 
good sorting algorithms for evolving data 
and to study how sensitive they are to 
parameters in the underlying evolving data model. 
Interestingly, our experiments provide empirical evidence
that quadratic-time sorting algorithms, including 
bubble sort, cocktail sort, and insertion sort, can outperform more 
sophisticated algorithms, such as quicksort and even the batched
parallel blocked quicksort algorithm of 
Anagnostopoulos~{\it et al.}~\cite{sort11}, in practice.
Furthermore, our results also show that 
even though these quadratic-time algorithms perform compare-and-swap 
operations only for
\emph{adjacent} elements at each time step, they are nevertheless robust 
to increasing the rate, $r$, at which random swaps occur in the underlying 
list for every comparison done by the algorithm. 
That is, our results show that these quadratic-time algorithms
are robust even in the face of 
an adversary who can perform many swaps for
each of an algorithm's comparisons. Moreover, this robustness
holds in spite of the fact
that, in such highly volatile situations,
each element is, on average, moved more often by random swaps 
than by the sorting algorithm's operations.

We also introduce the hot spot adversary and 
study sorting algorithms in the evolving data model 
with respect to this adversary. Our experiments provide evidence that
these sorting algorithms have similar robustness against the hot spot
adversary as they do against the uniform adversary.
Finally, we show that the constant factors in
the Kendall tau distances maintained by classic quadratic-time
sorting algorithms appear to be quite reasonable in practice.
Therefore, we feel our experimental results are arguably surprising, in
that they show the strengths of quadratic-time
algorithms in the evolving data model, in spite of 
the fact that these algorithms are much maligned 
because of their poor performance in the traditional
Knuthian model.

With respect to the organization of our paper,
\autoref{sec:prelim} formally introduces the evolving sorting model as well as the algorithms we consider in this paper. \autoref{sec:exps} describes the experiments we performed and presents our results.

\section{Preliminaries}\label{sec:prelim}

% Anagnostopoulos~{\it et al.}~\cite{sort11} formulated the problem of sorting evolving data. 
Let us begin by formally defining
the evolving data model for sorting,
based on the pioneering work of Anagnostopoulos~{\it et al.}~\cite{sort11}.
We assume that
there are $n$ distinct elements that belong to a total order relation, ``$<$''.
During each time step, a sorting algorithm is allowed to perform one comparison of a pair of elements and then 
an adversary is allowed to perform some random swaps
between adjacent elements in the true total order. In this paper,
we consider two types of adversaries:
\begin{enumerate}
\item
The \emph{uniform} adversary. 
This adversary performs 
a number, $r > 0$, of swaps between pairs of adjacent elements
in the true total order, where
these pairs are chosen uniformly at random and independently for each
of the $r$ swaps.
\item
The \emph{hot spot} adversary.
This adversary chooses an element, $x$, in the total order and then
randomly choose a direction, up or down.
The adversary then randomly chooses a bit, $b$. If this bit is $0$,
he swaps $x$ with its predecessor (resp., successor), depending on
the chosen direction, and repeats this process with a new random bit,
$b$. If $b=1$, he stops swapping $x$.
\end{enumerate}

We denote the state of the list the algorithm maintains at time $t$ 
with $\ell_t$ and the state of the unknown true ordering with $\ell'_t$.
If the time step is clear from the context or is the most current
step, then we may drop the subscript, $t$.
The main type of adversary that we consider is the same as that considered
in Anagnostopoulos~{\it et al.}~\cite{sort11}; namely, 
what we are calling
the uniform adversary.
Note that with this adversary,
after each comparison, a uniformly random adjacent pair of 
elements in $\ell'$ exchange positions and this process is repeated
independently for a total of $r>0$ swaps. 
We call each such change to $\ell'$ a \emph{swap} mutation. 
With respect to the hot spot adversary, we refer to the change made
to $\ell'$ as a \emph{hot spot} mutation, i.e.,
where an element is picked uniformly at random and flips an unbiased
coin to pick left or right, and then an unbiased coin is flipped
until it comes up heads and the element is swapped in the chosen
direction that many times.

Note that with either adversary,
a sorting algorithm cannot hope to correctly maintain the true
total order at every step, for at least the reason
that it has no knowledge of how the most recent mutation affected 
the underlying order. Instead, a sorting algorithm's goal is to
maintain a list of the $n$ elements that has a small {Kendall tau} 
distance, which counts the number of inversions\footnote{Recall that an
  \emph{inversion} is a pair of elements $u$ and $v$ such that $u$ comes before
  $v$ in a list but $u>v$.}
relative to the underlying total order.

We abuse the names of the classical sorting algorithms to refer to 
evolving sorting algorithms that repeatedly run that classical
algorithm. For instance, the insertion sort evolving sorting algorithm 
repeatedly runs the classical in-place insertion sort algorithm. 
We refer to each individual run of a classical sorting algorithm as 
a \emph{round}.

In this paper, we consider several classical sorting algorithms, which are
summarized in simplified form 
below (see~\cite{Cormen:2001,Goodrich:2014:ADA} for details 
and variations on these algorithms):
\begin{itemize}
\item \emph{Bubble sort}.
For $i=1,\ldots,n-1$, repeatedly compare the
elements at positions $i$ and $i+1$, swapping them if they
are out of order. Repeat $n-1$ times.

\item \emph{Cocktail sort}.
For $i=1,\ldots,n-1$, repeatedly compare the
elements at positions $i$ and $i+1$, swapping them if they
are out of order. Then do the same for $i=n-1,\ldots,1$. Repeat
$(n-2)/2$ times.

\item \emph{Insertion sort}. For $i=2,\ldots,n$, compare the element,
$x$, at position $i$ with its predecessor, swapping them if they are out of
order. Then repeat this process again with $x$ (at its new position)
until it reaches the beginning of the list or isn't swapped. Repeat
$n-2$ times.

\item \emph{Quicksort}. Randomly choose a \emph{pivot}, $x$, in the
list and divide the list inplace as the elements that are less than
or equal to $x$ and the elements that are greater than $x$.
Recursively process each sublist if it has at least two elements.
\end{itemize}

We also consider the following
algorithm due to Anagnostopoulos~{\it et al.}~\cite{sort11}:
\begin{itemize}
\item \emph{Block sort}.\footnote{The quicksorts of the entire 
   list guarantee that no element is 
   more than $O(\log n)$ positions from
   its proper location in the true sorted order. The quicksorts of the blocks account for elements of the list drifting away from their original positions.}
Divide the list into overlapping blocks of size
$O(\log n)$. Alternate between steps of quicksort on the entire list and steps of quicksort on each block.

%Repeat this process again, but with the blocks shifted so that the
%boundary of each block lands in the middle of a block in the previous
%phase.
%That is, the blocks overlap from one phase to the next, where each
%phase involes performing a quicksort of each block.
\end{itemize}

There are no theoretical results on the quality of bubble, cocktail, or
insertion sort in the evolving data model. 
Anagnostopoulos~{\it et al.}~\cite{sort11} showed that quicksort 
achieves $\Theta(n \log n)$ inversions w.h.p.~for any 
small constant, $r$, of swap mutations. 
Anagnostopoulos~{\it et al.}~\cite{sort11} also showed that block
sort achieves $\Theta(n \log \log n)$ inversions w.h.p.~for any small
constant, $r$, of swap mutations.

These algorithms can be classified in two ways.
First, they can be separated into the worst-case quadratic-time sorting 
algorithms (bubble sort, cocktail sort, and insertion sort) and the
more efficient algorithms (quicksort and block sort),
with respect to their performance in the Knuthian model.
Second, they can also be separated into two classes based on the types of comparisons they perform and how they update $\ell$. 
Bubble sort, cocktail sort, and insertion sort consist of 
compare-and-swap operations of adjacent elements in $\ell$ and 
additional bookkeeping, while quicksort and block sort perform comparisons 
of elements that are currently distant in $\ell$ and only 
update $\ell$ after the completion of some subroutines.

During the execution of these algorithms, the Kendall tau distance does not converge to a single number. For example, the batch update behavior of quicksort causes the Kendall tau distance to oscillate each time a round of quicksort finishes. Nevertheless, for all of the algorithms, the Kendall tau distance empirically reaches a final repetitive behavior. We call this the \emph{steady behavior} for the algorithm and judge algorithms by the average number of inversions once they reach their steady behavior. We call the time it takes an algorithm to reach its steady behavior its \emph{convergence time}. Every algorithm's empirical convergence time is at most $n^2$ steps.

\section{Experiments}\label{sec:exps}
The main goal of our experimental framework is to address the
following questions:

\begin{enumerate}
\item 
Do quadratic-time algorithms actually perform as well as or better than
quicksort variants on evolving data 
using swap mutations, e.g., for reasonable values of $n$?
\item 
What is the nature of the convergence of sorting algorithms on
evolving data, e.g., how quickly do they converge and 
how much do they oscillate once they converge?
\item 
How robust are the algorithms to increasing the value of $r$
for swap mutations?
\item
How much does an algorithm's convergence behavior and steady behavior
depend on the list's initial configuration
(e.g., randomly scrambled or initially sorted)?
\item 
How robust are the algorithms to a change in the mutation
behavior, such as with the hot spot adversary?
\item
What is the fraction of random swaps that actually improve Kendall
tau distance?
\end{enumerate}

We present results below that address each of these
questions.

\subsection{Experimental Setup}

We implemented the various algorithms that we study
in C++11 and our software is available on Github. \footnote{See code at \url{https://github.com/UC-Irvine-Theory/SortingEvolvingData}.}
Randomness for experiments was generated using the C++ default\_engine and 
in some cases using the C random engine.
In the evolving data model, each time step corresponds to one
comparison step in an algorithm, which is followed by a mutation
performed by an adversary. 
Therefore, our experiments measure time by the number of comparisons
performed.
Of course, all the algorithms we study are comparison-based; hence,
their number of comparisons correlates with their 
wall-clock runtimes.
To measure Kendall tau distances,
we sample the number of inversions 
every $n/20$ comparisons, where $n$ is the size of a list. 
We terminate a run after $n^2$ time steps and well after the algorithm
 has reached its steady behavior. 
The experiments primarily used random swap mutations;
hence, we omit mentioning the mutation type unless we are using hot spot mutations.

Anagnostopoulos~{\it et al.}~\cite{sort11} does not give an exact block size 
for their block sort algorithm except in terms of several unknown constants.
In our block sort implementation, the block size chosen for block sort is the first even number larger than $10 \ln n$ that divides $n$. Because all of the $n$ in our experiments are multiples of $1000$, the block size is guaranteed to be between $10\ln n$ and $100\ln n$.

% We are interested in how efficiently each algorithm behaves relative to how many comparisons have been executed, in a sense we are interested in the efficiency of these compares when handling evolving data
% For this reason our experiments compare the algorithms relative to the number of comparisons executed. 
% An unfortunate consequence is that there is an inconsistency between the algorithms because a large amount of work and book keeping can be done without executing a comparison. The clearest case is how many non-comparison operations are executed between a step of bubble sort and one of quicksort. In our experiments this doesn't translate to an advantage thus we choose to ignore this factor and focus on the comparisons.

\subsection{General Questions Regarding Convergence Behavior}

\begin{figure*}[!hbt]
\vspace*{-10pt}
\centering
\includegraphics[scale=0.5]{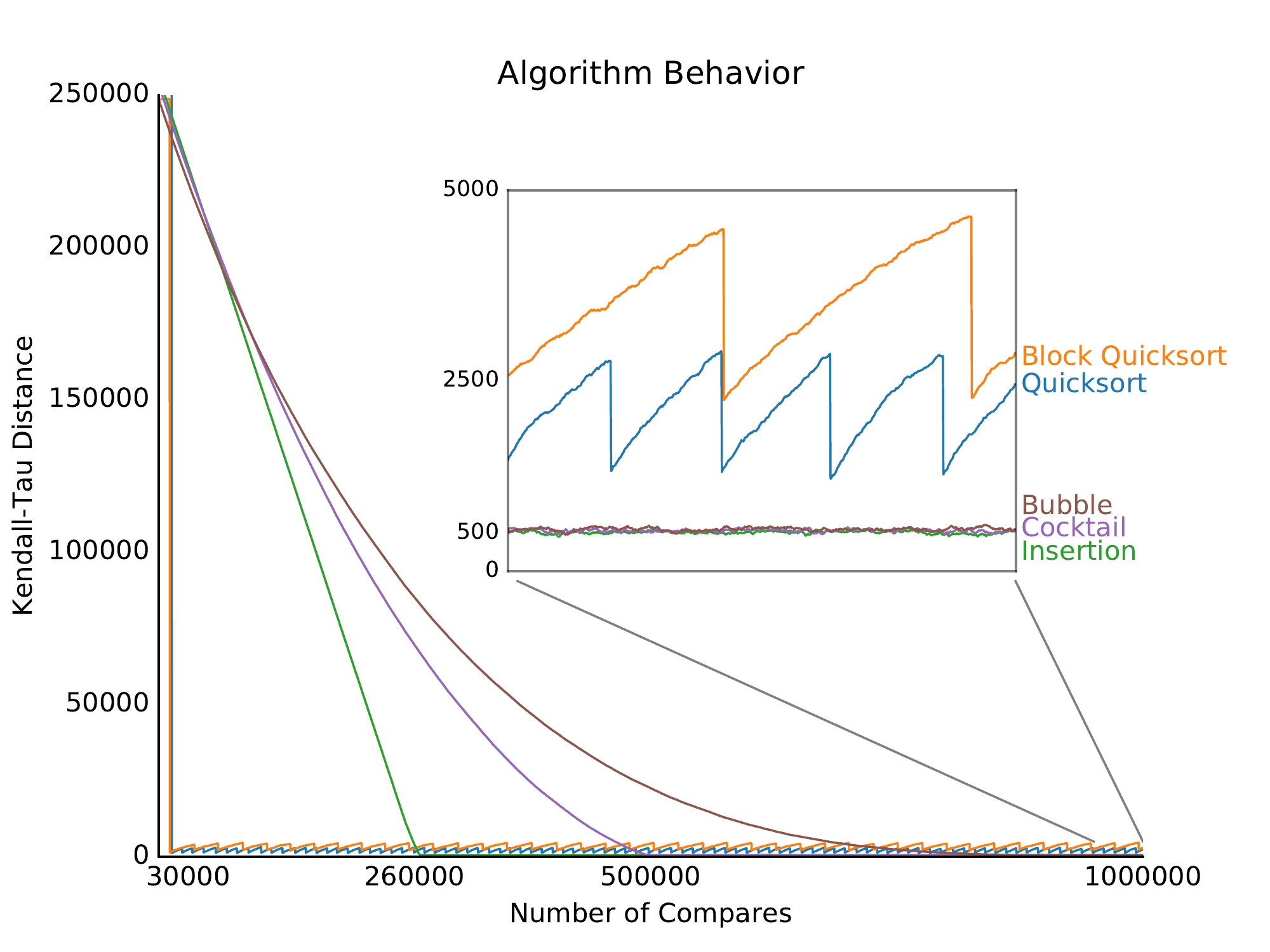}
\vspace*{-5pt}
\caption{Behavior of the algorithms starting from a shuffled list.
The plot shows Kendall tau distance as a function of an algorithm's
execution, i.e., number of comparisons, with random swap mutations
for $r=1$.
We also show an enlarged portion of the tail-end steady behaviors.
}\label{fig:algs}
\end{figure*}

We begin by empirically addressing the first two
questions listed above, which concern 
the general convergence behavior of the different algorithms.
\autoref{fig:algs} shows Kendall tau distance achieved by 
the various algorithms we studied as a function of the algorithm's 
execution time, against the uniform adversary 
(i.e., with random swap mutations), for the case when $r=1$ and $n = 1000$ starting from a uniformly shuffled list.

The quadratic time algorithms run continuous passes on the list 
and every time they find two elements in the wrong order they 
immediately swap them. 
That is, they are local in scope, at each step fixing inversions 
by swapping adjacent elements. 
They differ in their approach, but once each such algorithm
establishes a balance between the comparisons performed by the algorithm 
and the mutations performed by the uniform
adversary, their Kendall tau numbers don't differ significantly.  
These algorithms have very slow convergence because they only compare adjacent elements in the list and so fix at most one inversion with each comparison.
The Kendall tau behavior of the quicksort algorithms, on the other
hand, follow an oscillating pattern of increasing Kendall tau 
distance until a block (or recursive call) is finished,
at which point $\ell$ is quickly updated,
causing a large decrease in Kendall tau distance.

As can be seen, 
the convergence behavior of the algorithms can be classified into two groups.
The first group consists of the two quicksort variants, which very
quickly converge to steady behaviors that oscillate in a small range.
The second group consists of the quadratic-time algorithms, which
converge more slowly, but to much smaller Kendall 
tau distances and with no clear oscillating behavior.
More interestingly, the quadratic-time algorithms all converge to the
same tight range and this range of values is better than the wider
range of Kendall tau distances achieved by the quicksort algorithms.
Thus, our first experimental result already answers 
our main question, namely, that these quadratic-time algorithms appear
to be optimal for sorting evolving data and this behavior is actually 
superior in the limit to the quicksort variants.
% The first questions our experiments answer 
% there are two groups the batch algorithms that quickly
% converge to a stable behavior and the quadratic algorithms which
% converge more slowly but after that have less inversions.

Given enough time, however, all three quadratic-time algorithms
maintain a consistent Kendall tau distance in the limit.
Of the three quadratic-time algorithms,
the best performer is insertion sort, followed by cocktail sort and
then bubble sort.
% an order that is closest to the underlying distance. 
% The only exception is when the number of mutations after each compare is very large in which case quicksort may perform better. 
The worst algorithm in our first batch of experiments was block quicksort. This may be because $n=1000$ is too small for the theoretically proven $O(n \log \log n)$ Kendall tau distance to hold.

\subsection{Convergence Behavior as a Function of $r$}
Regardless of the categories, after a sufficient number of comparisons,
all the algorithms
empirically reach a \emph{steady behavior} where the distance 
between $\ell$ and $\ell'$ follows a cyclic pattern over time. 
This steady behavior depends on the algorithm, the list size, $n$,
and the number of random swaps, $r$, per comparison,
but it is visually consistent across many different runs and 
starting configurations.

\begin{figure*}[!hbt]
\vspace*{-2pt}
\centering
\includegraphics[scale=0.45]{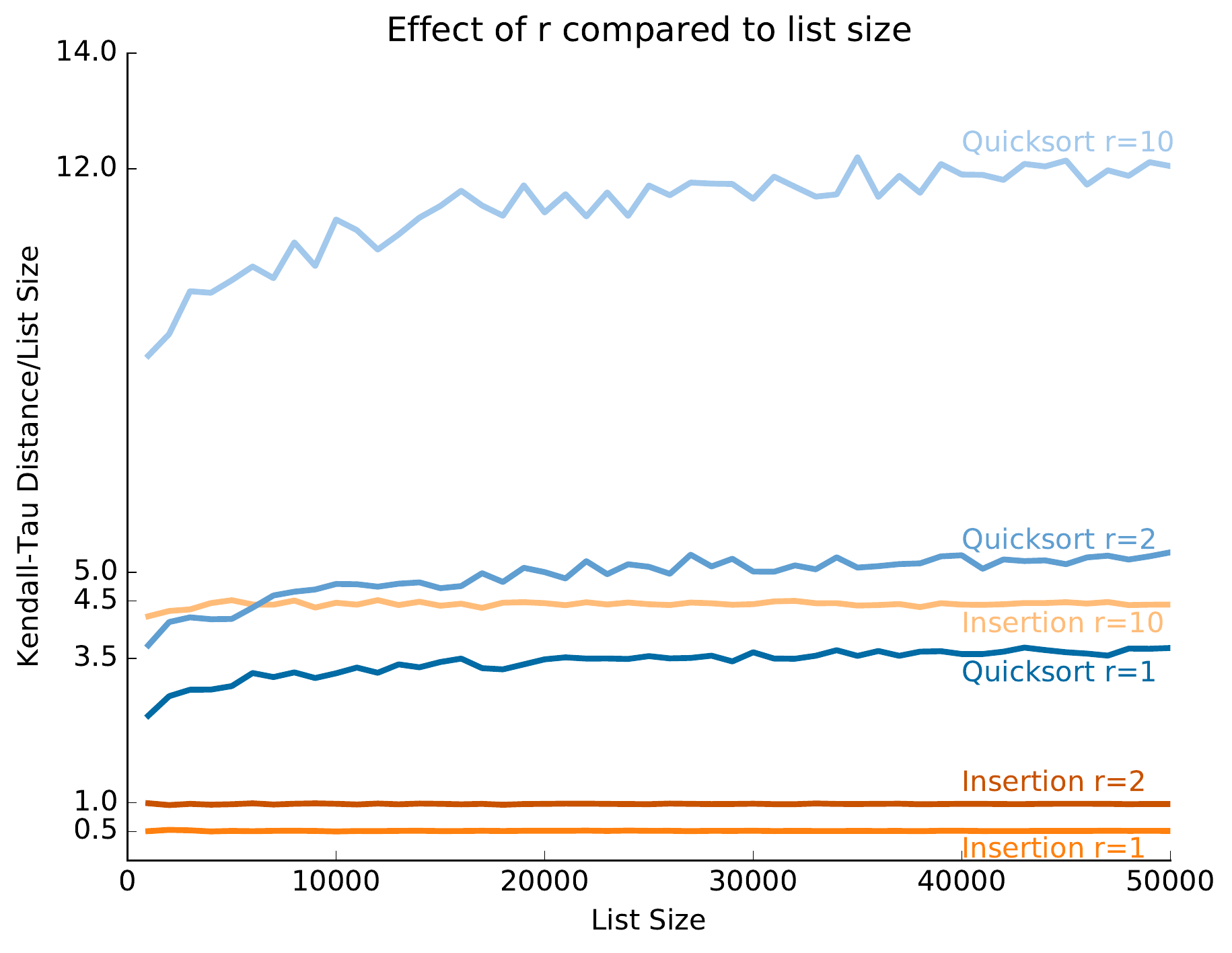}
\vspace*{-8pt}
\caption{Convergence ratios as a function of list size, $n$, and number of
random swaps, $r$, per comparison. 
% Shown are the results for insertion sort and quicksort. 
The vertical axis shows the ratio, $K/n$, where $K$ is 
the average Kendall tau value (number of inversions) in the steady behavior, 
and the horizontal axis shows the number of elements,~$n$. 
The curves show the behaviors of insertion sort and quicksort for $r\in\{1,2,10\}$.}
\label{fig:rVsSize}
\vspace*{-10pt}
\end{figure*}

Our next set of experiments, therefore, are concerned with studying convergence 
behavior as a function of $n$ and $r$.
We show in \autoref{fig:rVsSize} the convergence values comparing 
insertion sort and quicksort, as a ratio, $K/n$, of the steady-state Kendall tau
distance, $K$ (averaged across multiple samples once an algorithm
has reached its steady behavior),
and the list size, $n$.

As can be seen from the plots, for these values of $r$, insertion sort
consistently beats quicksort in terms of the Kendall tau distance it achieves,
and this behavior is surprisingly robust even as $n$ grows.
Moreover, all of the quadratic-time algorithms that we studied achieved similar
Kendall-tau-to-size ratios that were consistently competitive with both quicksort
and block sort.
In the appendix, we give a table showing specific ratio values for these algorithms
for a variety of chosen values for $r$.
Our results show that for values of $r$ larger than $50$, the quicksort variants
tend to perform better than the quadratic-time algorithms, but the quadratic time
algorithms nevertheless still converge to reasonable ratios.
We show in \autoref{table:convTable}, which is also given 
in the appendix, specific convergence values for different
values of $r$, from $0$ to $10$, for insertion sort and quicksort.

\subsection{Starting Configurations}

The quadratic time algorithms all approach their steady behavior 
in a similar manner, namely, at an approximately constant rate attenuated by $r$.
Thus,
we also empirically investigated how long each algorithm requires to reach a 
steady behavior starting from a variety of different start configurations. 

\begin{figure*}[!hbt]
\centering
\includegraphics[scale=0.5]{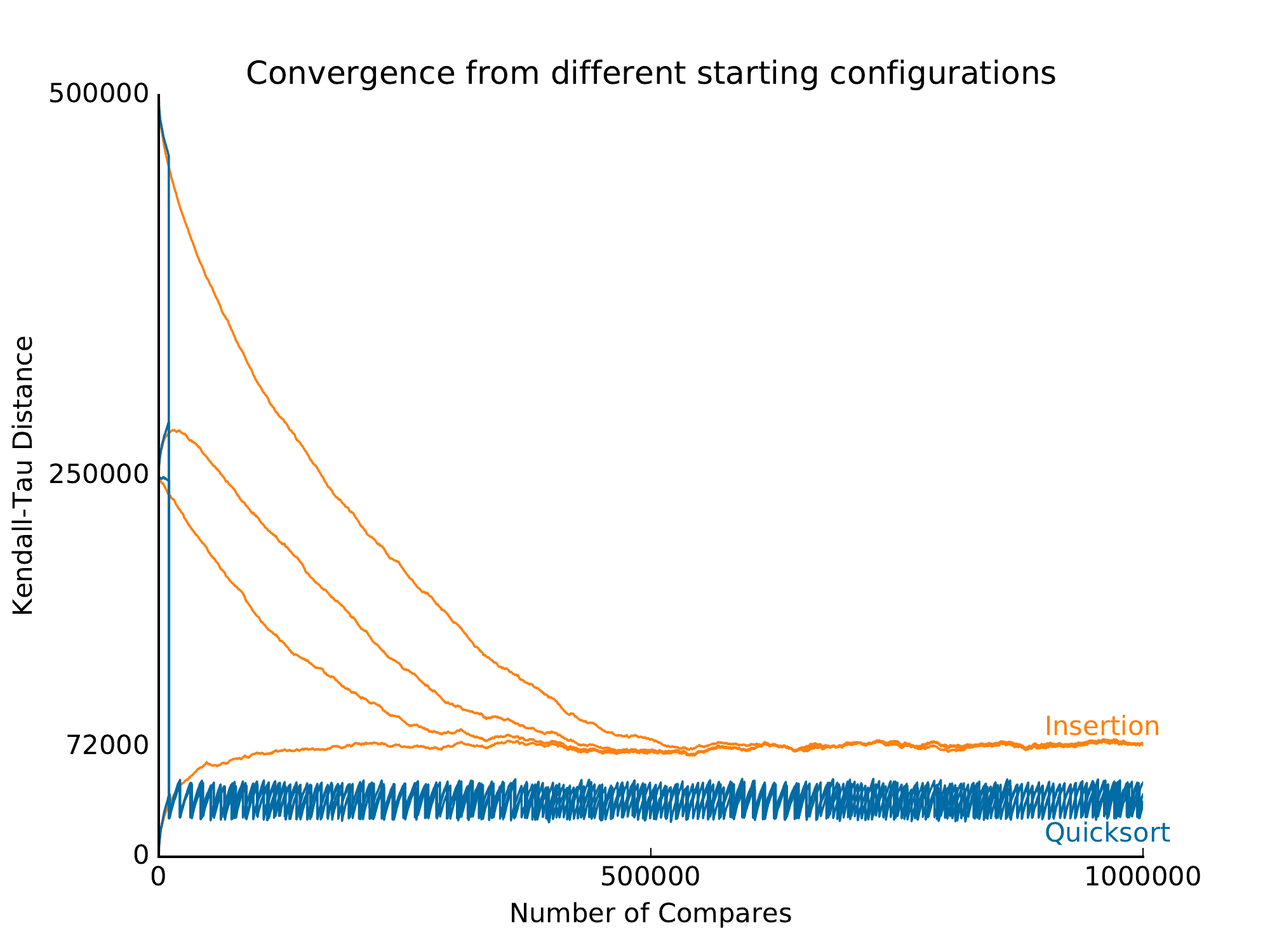}
\caption{Even for a large $r = 256$ over only 1000 elements the
algorithms quickly converge to a steady behavior.
The plots show the Kendal tau distances achieved for insertion sort
and quicksort for four starting configurations of increasing complexity:
(i) initially sorted, (ii) random shuffle, (iii) half cyclical shift of a sorted
list, and (iv) a list in reverse order.}\label{fig:startConfig}
\end{figure*}

Because both quicksort and block sort begin with a quicksort round, 
their Kendall tau distance drops quickly after $O(n\log n)$ comparisons.
The other algorithms require a number of comparisons proportional to the initial distance from the steady-state value. 
For example, see \autoref{table:rTable} in the appendix, 
which shows that insertion sort's steady behavior when $r=1$ 
is roughly $n/2$ inversions. 
When the initial state of a list has $I$ inversions, 
the number of comparisons that insertion sort requires to reach 
the steady behavior is approximately $4|I-n/2|$. 
Thus,
if the list is initially sorted, insertion sort will take 
approximately $2n$ comparisons to reach its steady behavior.
Moreover, increasing $r$ does not seem to affect the convergence rate significantly. 
\autoref{fig:startConfig}
shows a plot illustrating the convergence behavior of insertion sort and quicksort
for a variety of different starting configurations.

Our experiments show that, as expected, 
the convergence of insertion sort is sensitive to the starting configuration,
whereas quicksort is not.
Primed with this knowledge,
these results justify our starting from 
a sorted configuration in our subsequent experiments, so as to explore convergence
values without having to wait as long to see the steady behavior.
Still, it is surprising that the quadratic algorithms converge at all for $r=256$
in these experiments, 
since for each inversion fixed by a quadratic algorithm 
the adversary gets to swap 256 pairs in a list of 1000 elements.

In general,
these early
experiments show that, after converging, 
the quadratic time algorithms perform significantly better 
than the more efficient algorithms for reasonable values of $r$ and that they are
competitive with the quicksort values even for larger values of $r$.
But in an initial state with many inversions, 
the more efficient algorithms require fewer compares to quickly 
reach a steady behavior. Thus,
to optimize convergence time, it is best to run an initial
round of quicksort and then switch to repeated rounds
consisting of one of the quadratic-time algorithms.

\subsection{Hot spot mutations}

\begin{figure*}[!hbt]
\vspace*{-10pt}
\centering
\includegraphics[scale=0.48]{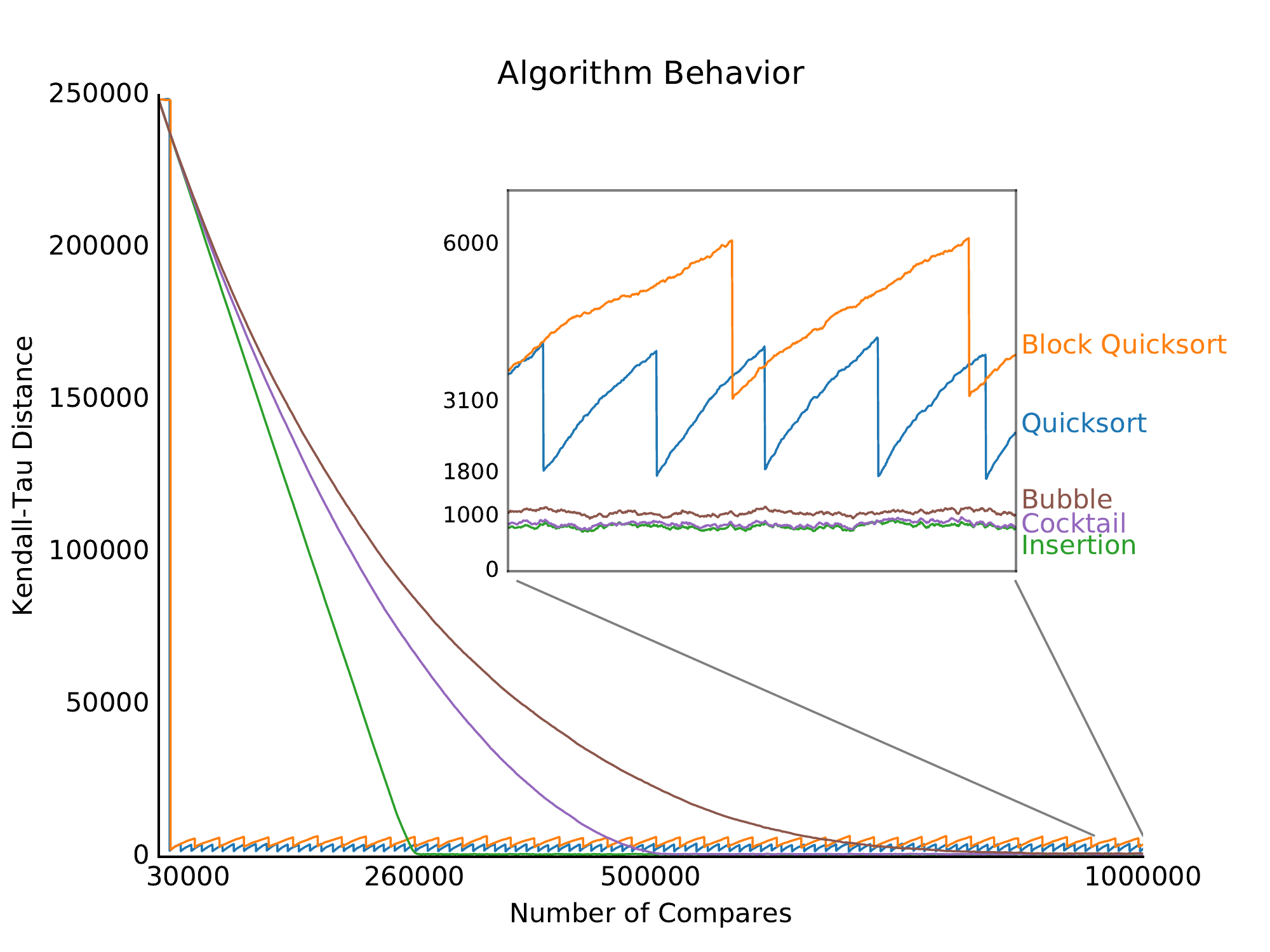}
\vspace*{-5pt}
\caption{Impact of hotspot mutations. 
% Hotspot mutations impact quadratic algorithms twice as much as uniform mutations.
We plot Kendall tau distance as a function of an algorithm's
number of comparisons.
We also show an enlargement of the tail-end steady behaviors.
}\label{fig:hot}
\end{figure*}

Recall that hot spot mutations simulate an environment in which, instead of 
a pair of elements swapping with each other, an
element changes its rank based on a geometric distribution.
\autoref{fig:hot} 
shows the convergence behavior of the various algorithms 
against the hot spot adversary.
Comparing \autoref{fig:hot} to \autoref{fig:algs}, we see 
that the quadratic algorithms are twice as affected 
by hot spot mutations as by uniform mutations, although the total number 
of adversarial swaps is the same in expectation. 

A possible explanation for this behavior is that
a large change in rank of a single element in a hot spot mutation
can block a large number of other
inversions from being fixed during a round of a quadratic algorithm.
For example, in insertion sort, during a round, the element at each 
position, $0 \le i \le n$, is expected to be moved to its 
correct position with respect to $0,\dots,i-1$ in $l$. 
To do so, an element $x$ is swapped left until it reaches a smaller element $y$. 
But
if $y$ has mutated to become smaller, then all the elements left of $y$ 
in $l$, which remain larger than $x$, will stay inverted with respect to $x$ 
until the end of the round (unless some other mutation fixes some of them). 
Bubble and cocktail sort have a similar problem. 
An element $x$ which mutates can block other local, smaller inversions involving 
elements in between $x$'s starting and ending position. When these inversions
are not be fixed, hot spot mutations make each pass of these algorithms coarser.
Batch algorithms on the other hand are not affected as strongly by hot
spot mutations, because their behavior depends on non-local factors 
such as pivot selection and the location at which the list was partitioned. 
Therefore, the movement of a single element has a smaller effect on their behavior.
Thus, we find it even more surprising that the quadratic algorithms
still beat the quicksort variants even for the hot spot adversary (albeit
by a lesser degree than the amount they beat the quicksort variants
for the uniform adversary).

\subsection{Beneficial Swaps Performed by an Adversary}

\begin{figure*}[!hbt]
\centering
\includegraphics[scale=0.5]{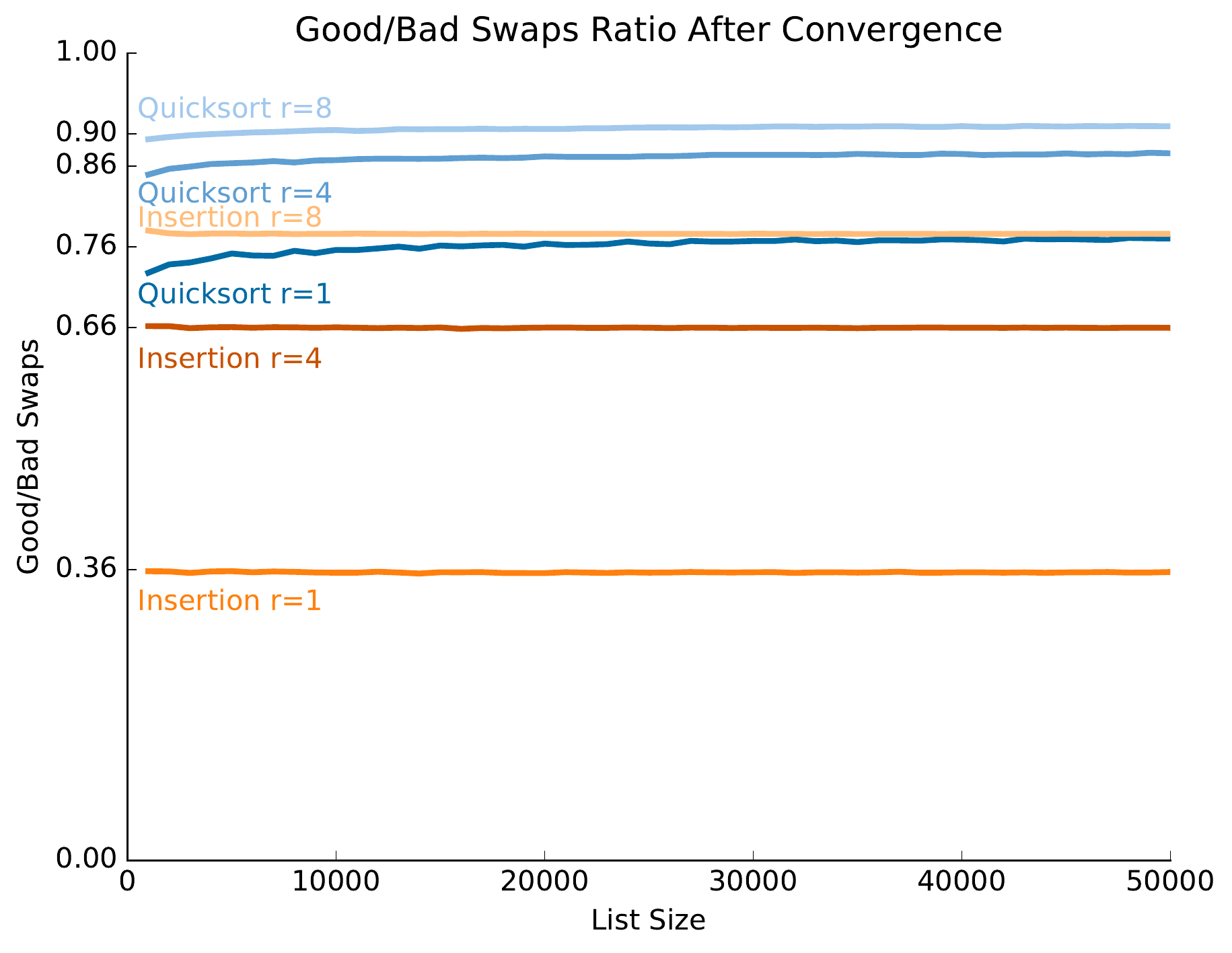}
\caption{As $r$ increases, the number of inversions increases, but so does the number of beneficial mutations.}\label{fig:swapRatio}
\end{figure*}

Note that our quadratic-time algorithms compare only adjacent elements, so they can only fix one inversion at each step. Therefore, they will not reach a steady state until the random swaps fix inversions almost as often as they introduce inversions. \autoref{fig:swapRatio} shows that the proportion of good swaps (those that fix inversions) to bad swaps (those that introduce inversions) approaches $1$ as $r$ increases.
This behavior might be useful to exploit, therefore, in future work that would provide theoretical guarantees for the performance of bubble sort and cocktail sort in the evolving data model.

\section{Conclusion}
We have given an experimental evaluation of sorting in the evolving
data model. Our experiments provide empirical evidence that,
in this model,
quadratic-time algorithms can be superior to algorithms
that are otherwise faster 
in the traditional Knuthian model.
We have also studied a number of additional questions regarding 
sorting in the evolving data model.
Given the surprising nature of many of our results,
it would be interesting
in the future to empirically study algorithms for other problems besides sorting in 
the evolving data model.
Alternatively, it would also be interesting to provide theoretical analyses for
some of the experimental phenomena that we observed for sorting in the 
evolving data model, such as the performance of algorithms against 
the hot spot adversary.

\bibliographystyle{plainurl}
\bibliography{refs}

\begin{appendix}
\section{Detailed Convergence Behavior}
In this appendix, we
provide detailed convergence results. 

In \autoref{table:convTable}, we show specific convergence rates between
insertion sort and quicksort, for a list of size 1000.
As can be seen, this table highlights the expected result that quicksort reaches
its steady behavior much more quickly than insertion sort.
Thus, this table provides empirical evidence supporting a hybrid algorithm
where one first performs a quicksort round and then repeatedly 
performs insertion sort rounds after that.

\begin{table}[!hbt]
\centering
\begin{tabular}{l | c c }
 $r$ & Insertion & Quicksort \\
 \hline
 0 & 500000 & 12000 \\
 1 & 510000 & 16000 \\
 2 & 513000 & 16000 \\
 3 & 516000 & 15000 \\
 4 & 516000 & 16000 \\
 5 & 521000 & 16000 \\
 6 & 523000 & 16000 \\
 7 & 521000 & 17000 \\
 8 & 525000 & 17000 \\
 9 & 524000 & 17000 \\
10 & 527000 & 16000 \\
\end{tabular}
\caption{Number of comparisons needed to converge to steady behavior for different values of $r$.}
\label{table:convTable}
\end{table}

In \autoref{table:rTable},
we show the ratios of the number of inversions to list size for various
values of $r$, with respect to the uniform adversary and multiple algorithms.
Note that the ratios grow slowly as a function of $r$ and that the quadratic time
algorithms are better than the quicksort variants for values of $r$ up to around
50.  After that threshold, standard quicksort tends to perform better than the
quadratic-time algorithms, but the quadratic-time algorithms nevertheless 
still converge and perform reasonably well.
Interestingly, the quadratic algorithms still beat block sort even for these
large values of $r$.

\begin{table*}[!hbt]
\centering
\begin{tabular}{l | c c c c c}
 $r$ & Insertion & Cocktail & Bubble & Quicksort & Block Quicksort \\
 \hline
  1 &  0.51 &  0.54 &  0.54 &  2.17 &  4.03 \\
  2 &  0.98 &  0.98 &  1.13 &  3.40 &  5.78 \\
  3 &  1.45 &  1.42 &  1.64 &  4.24 &  7.19 \\
  4 &  1.84 &  1.76 &  2.17 &  4.51 &  8.58 \\
  5 &  2.28 &  2.04 &  2.69 &  5.03 &  9.85 \\
  6 &  2.72 &  2.46 &  3.05 &  5.83 & 10.11 \\
  7 &  3.16 &  2.83 &  3.40 &  6.62 & 11.39 \\
  8 &  3.49 &  3.20 &  3.89 &  7.15 & 12.06 \\
  9 &  4.03 &  3.63 &  4.50 &  7.04 & 12.74 \\
 10 &  4.37 &  3.87 &  4.96 &  7.45 & 14.09 \\
 11 &  4.64 &  4.09 &  5.58 &  7.44 & 14.60 \\
 12 &  5.07 &  4.61 &  5.79 &  8.12 & 15.91 \\
 13 &  5.32 &  4.92 &  6.17 &  7.96 & 15.89 \\
 14 &  5.91 &  5.14 &  6.73 &  9.35 & 16.36 \\
 15 &  6.21 &  5.76 &  6.94 &  9.75 & 17.55 \\
 16 &  6.52 &  5.98 &  7.33 &  9.91 & 17.54 \\
 17 &  7.06 &  6.05 &  7.74 & 10.03 & 18.21 \\
 18 &  7.56 &  6.43 &  8.13 & 10.02 & 18.59 \\
 19 &  7.79 &  6.94 &  8.56 & 10.38 & 19.73 \\
 20 &  8.25 &  7.51 &  8.68 & 10.89 & 20.93 \\
 $\dots$ &  &       &       &       &       \\
 40 & 14.98 & 13.53 & 17.12 & 15.05 & 25.18 \\
 41 & 15.32 & 14.27 & 17.86 & 15.19 & 25.24 \\
 42 & 15.79 & 14.11 & 17.77 & 15.46 & 25.00 \\
 43 & 16.26 & 14.36 & 17.79 & 15.34 & 26.85 \\
 44 & 16.42 & 14.74 & 18.05 & 15.79 & 26.39 \\
 45 & 16.45 & 15.11 & 18.73 & 15.60 & 27.81 \\
 46 & 17.09 & 15.40 & 19.31 & 16.09 & 27.47 \\
 47 & 17.37 & 15.70 & 19.73 & 16.36 & 27.32 \\
 48 & 17.42 & 16.02 & 19.97 & 16.21 & 28.55 \\
 49 & 17.87 & 16.22 & 20.08 & 16.46 & 28.08 \\
 50 & 18.55 & 16.57 & 20.66 & 16.72 & 29.23 \\
 $\dots$ &  &       &       &       &       \\
100 & 32.67 & 30.36 & 35.18 & 23.83 & 43.18 \\
256 & 65.20 & 61.30 & 66.20 & 38.10 & 74.53 \\

\end{tabular}
\caption{Ratio of inversions relative to list size for different values of $r$}
\label{table:rTable}
\end{table*}

\end{appendix}

\end{document}